Absence of Observed Unspeakably Large Black Holes Tells Us the Curvature of Space

*David Alan Johnson, Dept. of Philosophy, Yeshiva University*

In this paper I argue for the initial conclusion that (using 'trillion' in the American sense, of $10^{12}$) there is less than one chance in a million trillion that space is Euclidean (as will become clear below, this can be greatly reduced: to barely one chance in a hundred thousand trillion trillion; and perhaps ultimately to the nether region of $10^{-357000}$), and for the initial conclusion that there are less than two chances in a million trillion that space is not spherical (this can be similarly greatly reduced). But in fact so great is the probability against the hypothesis that space is *hyperbolic* that this hypothesis can simply be ruled out (by a black-hole argument; and, independently, by a simple proof making no appeal to black holes, but only to the ratio of volumes in hyperbolic space of Hubble volumes from different times). In the last section I argue for what will seem a strange conclusion about the things around us. That final argument, in its initial formulation, makes use (for convenience) of some surprising results about black holes, broached in the second section; chiefly the implication, from space being Euclidean, of the existence of black holes of unspeakable size: Hubble black holes.

Let me note that when I say "Euclidean" (or "spatially flat") I mean the supposition that space is *precisely* Euclidean ($k = 0$).

Throughout this work I use units in which $c = \mathrm{G} = h/2\pi = k_{\mathrm{B}}$ (Boltzmann's constant) $= 1$, though I always display $c$ and G. Thus mass density = energy density. It includes all contributions to mass-energy, such as the kinetic energies of particles.

I. Black-Hole Preliminaries

For the purpose of setting out some basic matters, I shall initially make the simplifying assumption that the black holes we are dealing with are overall electrically neutral, and non-rotating. Thus we will describe the spacetime outside of a spherically symmetric mass distribution by the Schwarzschild metric (refs [1]; [2], p. 268):

(1)     $ds^2 = -[1 - (r_S/r)]\,c^2 dt^2 + dr^2/[1 - (r_S/r)] + r^2\,(d\theta^2 + \sin^2\theta d\varphi^2)$

(in Schwarzschild coordinates $t$, $r$, $\theta$, $\varphi$) where

(2)     $r_S = 2GM/c^2$

is the Schwarzschild radius for the mass M.

I will assume that the metric inside a spherically symmetric mass distribution, in the *process of becoming*, or as a *candidate for being*, a Schwarzschild black hole, may be approximated as a region of a Friedmann-Lemaitre-Robertson-Walker (FLRW) spacetime  (as in the Oppenheimer-Snyder solution to spherical collapse to a Schwarzschild black hole; see refs [3]; [4], pp. 90-92; [5], pp. 18-22), with a suitable spherically symmetric spacetime slicing, so that the apparent horizon may be identified in the usual way. In a FLRW spacetime, the radius $R_{AH}$, measured from a point treated as the center of a sphere, to the sphere's *apparent horizon* is at any time $t$ (ref [6], p. 3):

(3)     $R_{AH} = c/[H_t^2 + (kc^2/a_t^2)]^{1/2}$

where $H_t$ is the (spatially constant) value at $t$ of the Hubble parameter (measuring the expansion rate at $t$ of the FLRW spacetime), $a_t$ is the scale factor (measuring at $t$ how much the spacetime has expanded; $H_t =_{df} (da_t/dt)/a_t)$, and $k$ is the Riemann curvature constant for the spacetime. For a spatially flat ($k = 0$) FLRW spacetime, $R_{AH}$ is thus $c/H_t$.

The interior metric for the candidate-black-hole region is a (spherical-space FLRW) line element derivable from the line element (e.g., (13) below) for a FLRW spacetime. The exterior metric is the Schwarzschild metric (1) above. That these two metrics (an FLRW metric for *inside* the region that is to become a black hole; the Schwarzschild metric for the spacetime *outside*) "can be joined smoothly at their common boundary" (the boundary of the spherically symmetric mass distribution) is well known (ref [4], p. 92).

As Misner, Thorne, and Wheeler put it beautifully in ref [7], p. 130: "spacetime contains a flowing river of 4-momentum." In general relativity, the stress-energy tensor captures all contributions to this river: mass-energy density, components of momentum density, components of energy flux, and components of momentum flux. In the Schwarzschild solution to the Einstein field equations, it is idealized that, outside the spherically symmetric mass distribution, the stress-energy tensor vanishes (the river doesn't flow): it is a solution (indeed, the unique solution) for vacuum outside a spherically symmetric mass source. (The phrase "spherically symmetric" refers to a certain constraint on the form of the metric tensor, which then yields the Schwarzschild metric. See refs [8], pp. 12-13, 638-639; [9], pp. 287-289. Henceforth I will just say "spherical mass distribution.") The Oppenheimer-Snyder solution to the Einstein field equations shows that a spherical mass distribution M may (for the purpose of assessing whether it becomes a black hole) be treated thus as though the exterior universe were literally a vacuum (which, in general relativity, implies that we are treating the spherical mass distribution as though it accrues no further mass). The current (or calculated) mass $M_c$ of the spherical mass distribution may thus be treated as the mass M in the Schwarzschild metric. As we explain below, in a Schwarzschild spacetime, if the matter-radius $r$ of M is $\leq r_S$ then M is a black hole, and $r_S$ is the radius to the event horizon of the black hole. But $M_c$, which actually inhabits a larger FLRW spacetime (we make the standard assumption that our universe on large scales is approximated by a FLRW spacetime), in which the exterior universe is not a vacuum, will certainly accrue further mass. And here we must mention a subtlety about black holes. The precise location at any given time of the event horizon—a null hypersurface; that is, a normal-vector-null-at-every-point $(2 + 1)$-dimensional ("a spatial 2-surface propagating in time"; ref [5], p. 9) submanifold of the 4-dimensional spacetime manifold; generated by certain null geodesics; that is, by certain (see ref [4], p. 203, for the general requirements) of the possible worldlines of things that always travel at $c$—depends on how much material ever falls into the black hole in the future history of the universe. (In the case of a black hole into which there is matter infall, this is because the null geodesics that generate the event horizon are the ones that "undergo just the right amount of focusing [by the infalling matter], so that after encountering the last of the infalling matter, their expansion goes to zero…The black hole's final state must be known before the horizon's null generators can be identified"; ref [4], p. 175; cf. ref [8], pp. 173-175.) It is determined, not locally, but by the global structure of spacetime, and thus the event horizon is often referred to as being *teleological*. So the "Schwarzschild radius" $r_S$ we get from the Oppenheimer-Snyder solution by treating $M_c$ as the M in the Schwarzschild metric is in fact not the radius to the event horizon of the black hole that $M_c$ is *within*. We will call $r_{PS} = 2GM_c/c^2$ the *pseudo-Schwarzschild radius*. Where $r_c$ is the current (or calculated) radius of the spherical mass distribution $M_c$, we know by the Oppenheimer-Snyder solution that if $r_c \leq r_{PS}$ then $M_c$ is *within a black hole* (for $M_c$ is treated as the M of the Schwarzschild metric, with matter-radius $r = r_c$, and if $r \leq r_S$ then M is a black hole). But assuming that $M_c$ accrues further mass, the precise identification at any given time of the radius $r_{EH}$ to the event horizon, even when the spherical mass distribution is precisely $M_c$, is a matter of notorious difficulty (as in a Vaidya spacetime; see ref [4], pp. 173-175). Here, since we will be concerned only with minimum and maximum bounds of various kinds, and sometimes with mere possibilities, we may deal with the difficulty in the following way:

(i) In those cases where the object or region spoken of (the black-hole-candidate) is merely imaginary and illustrative, I will simply assume that, if the object or region is a black hole, then so little further material ever falls into the black hole (just neutrinos, photons, and the like) that there is no vast difference between the pseudo-Schwarzschild radius $r_{PS}$ and the radius $r_{EH}$ to the event horizon, for the stipulated current mass of the object or within the region; I will assume, rather arbitrarily, that they do not differ by more than an order of magnitude. It will be apparent that nothing hangs on the choice.

(ii) In those cases where the object or region spoken of is not imaginary, I will do the calculation for the pseudo-Schwarzschild radius and simply note that the point being made is not affected if the radius $r_{EH}$ to the event horizon is larger, vastly or otherwise.

Focusing now again just on the Schwarzschild metric (1) and the Schwarzschild radius (2), consider a spherical mass distribution M with matter-radius $r$, so that all of the mass M is within the sphere whose radius is $r$. Then if $r \leq r_S$, the region enclosed by the sphere is a black hole. Why do I say "$r \leq r_S$", rather than "$r < r_S$"? After all, in (1) the singularity at $r = r_S$ is well known to be a mere coordinate singularity, and (unlike the one at $r = 0$) can be eliminated by transforming from Schwarzschild coordinates to Eddington-Finkelstein coordinates, or to Kruskal-Szekeres coordinates. However, "$\leq$" is correct. When $r_S/r$ reaches unity, so that M is "just within" its Schwarzschild radius, then the region enclosed by the sphere of radius $r_S$ is a black hole. (For then the sphere is a *boundary* separating matter and vacuum, and the Schwarzschild metric is applicable, and the metric's implications, noted below, take effect.) See Walecka (ref [10], pp. 199, 201, 214-215). Compare Steane's remark (ref [2], p. 290) that "[w]hen the gravitating body does not extend beyond its own Schwarzschild radius" then a black hole is formed. Compare Rindler's well-known example (ref [11]; ref [2], p. 290), of a hypothetical spherical galaxy (under the idealization that its exterior is a vacuum) which collapses to a point at which, although the stars are still far apart, the galaxy's radius is *equal* to the Schwarzschild radius of its mass, and the galaxy is become a black hole.

Why does a spherical mass distribution become a black hole when it shrinks to within the Schwarzschild radius of its mass? Some of the mass may or may not be distributed among literal "point" particles, literally in the sphere itself (though of course they would still be within the volume of the sphere). If there are such, forget about them for the moment, and just focus on the particles which are within the volume of the sphere but not in literally the sphere itself, so that for them $r < r_S$. Looking at (1), we notice that (as is standardly pointed out; cf. ref [2], p 292, ref [12], p. 553) for $r < r_S$ the coefficient in front of $dt^2$ is positive, and that in front of $dr^2$ is negative; that is, for $r < r_S$, time becomes spatial, and the $r$-component of space becomes time. It can be shown that for objects at $r < r_S$ motion "forward in time" then inevitably becomes motion toward $r = 0$, past which none can go since that would require moving "backward in time" (ref [2], p. 294). That is why the singularity at $r = 0$ becomes a point of infinite density. And that is why nothing within the volume of the sphere can escape to beyond the event horizon. (Any "point" particles literally in the sphere itself remain in it if they travel at $c$ directly away from the center; otherwise they too are doomed to reach $r = 0$.) That is why the sphere is an "event horizon"; a "point of no return." That is why the black hole is black. Let me quote Steane's remarks from ref [2], pp. 292-293:

Within the horizon something important happens to the Schwarzschild metric: the coefficient in front of d$t^2$ becomes positive, and that in front of d$r^2$ becomes negative. Therefore, intervals in the $t$ direction are space-like and intervals in the $r$ direction are time-like (the horizon itself is null). In short, despite the letter, $t$ now represents a spatial quantity and $r$ represents time. Particle worldlines remain time-like (after all nothing special is happening at the horizon, as Eddington-Finkelstein coordinates have taught us), but the central singularity is still at $r = 0$. The conclusion is that motion forward in time *is* motion towards smaller $r$. An object entering the horizon is carried down to $r = 0$ just as surely as you and I are carried into next week….Inside the horizon (or perhaps we should say *after* the horizon) the light-cones, and therefore all time-like intervals, and therefore all particle worldlines, tip over towards $r = 0$. It follows that once a star or other body manages to get completely inside its own Schwarzschild radius, it must collapse all the way to $r = 0$. No opposing force can be strong enough to prevent it.

There is no constraint on how (finitely) *small or large* the spherical mass distribution may be. We may take a spherical mass distribution on the order of Planck-scale lengths; or an "Earth" that is somehow compressed within a radius of .00887 meters; or a collapsing star; or Rindler's spherical galaxy; or a spherical mass distribution the size of the observable universe. Where charge and angular momentum may be ignored, there is no objection in principle to the application of the Schwarzschild metric to the spacetime outside any of these. All that is then necessary, for a spherical mass distribution to be a black hole, is that all of the mass is concentrated within, even "just within," the volume of a sphere whose radius is less than or equal to the Schwarzschild radius of the mass; that is, every bit of the mass is located at some $r \le r_S$.

In this paper I shall assume that, for the purpose of assessing whether a mass distribution enclosed in a sphere of radius $r'$ becomes a black hole—whether or not the mass distribution is *itself spherical* (i.e., spherically symmetric), and whether or not it has an electromagnetic field—it is as though we may apply an Oppenheimer-Snyder style solution, so that we have the following principle, which I will take to be uncontroversial: If the current mass $M_c$ enclosed by a sphere of radius $r'$ is such that $r' \le r_{PS}$ then the region enclosed by the sphere is within a black hole. Here I follow the practice of astrophysicists, when they infer that a celestial object or region "is a black hole," based simply on its calculated mass and matter-radius. Liddle and Loveday say of Sagittarius A*, at the heart of our galaxy: "Only a black hole can contain that much mass in such a small volume of space" (ref [13], p. 295). Harwit says: "A black hole is formed when an aggregate of mass $M$ is confined within a radius $R = 2MG/c^2$, where $G$ is the gravitational constant and $c$ is the speed of light" (ref [14], p. 41). Frolov and Novikov say (ref [8], p. 3):

A *black hole* is, by definition, a region in spacetime in which the gravitational field is so strong that it precludes even light from escaping to infinity.

A black hole is formed when a body of mass $M$ contracts to a size less than the so-called *gravitational radius* $r_g = 2GM/c^2$ ($G$ is Newton's gravitational constant, and $c$ is the speed of light). The velocity required to leave the boundary of the black hole and move away to infinity (the escape velocity) equals the speed of light. One easily concludes then that neither signals nor particles can escape from the region inside the black hole since the speed of light is the limiting propagation velocity for physical signals. This conclusion is of absolute nature in Einstein's theory of gravitation because the gravitational interaction is universal. The role of gravitational charge is played by mass whose value is proportional to the total energy of the system. Hence, all objects with nonzero energy [e.g., even photons] participate in the gravitational interaction.

(In the literature it is not uncommon to say "less than" in the sense of *within*; that is, "less than or equal to." That this is the intention here is clear from the remark that the escape velocity from "the boundary of the black hole" equals the speed of light. The authors show [pp. 40-42] that, at least in the case of the Schwarzschild spacetime, the relativistic escape velocity coincides with the Newtonian value of $(2GM/R)^{1/2}$ where R is the radius of the sphere enclosing mass M. So they indeed mean that the boundary is at the gravitational radius [= the pseudo-Schwarzschild radius]; since obviously $c = (2GM/R)^{1/2}$ if and only if $R = 2GM/c^2$.)

Sir Roger Penrose describes "the standard picture of collapse to a black hole…based on an assumption that no quantum-mechanical principles intervene to change the nature of space-time from that which is classically described by Einstein's general relativity," and based on the "assumption that *cosmic censorship*, in some form, holds true" (so that the singularity is "'shielded' from view by an *absolute event horizon*"), in the following way: "for bodies of too large a mass, concentrated in too small a volume, unstoppable collapse will ensue, leading to a *singularity* in the very structure of space-time…All that is required is for sufficient mass to fall into a small enough region" (ref [15], pp. 103-104). If "too much mass…concentrated in too small a volume" (p. 105) means that the mass is concentrated within a sphere whose radius is the pseudo-Schwarzschild radius of the mass, this may seem a hard thing to be wrought: perhaps only in the collapse of very massive stars, or when in "the central regions of a large galaxy…the required concentration could occur with the stars in the region still…separated from each other" (p. 104), or in peculiar microphysical circumstances. But in certain spatial geometries it is startlingly easy. In a Euclidean space there are very large regions which, if born with a mass density that is not below average, are born as black holes. They are called Hubble volumes.

However, we must address the matter that, in the Oppenheimer-Snyder solution, the exterior metric is the Schwarzschild metric; there is thus the assumption of spherical symmetry; and of the absence of an electromagnetic field (since it is an exterior vacuum solution). What if the mass distribution deviates from spherical symmetry; because it is rotating, or because, at some distance from the center, mass density is not constant at that distance? What if the mass distribution has an electromagnetic field? Can these things affect whether a black hole forms? Yes: because, in cases of gravitational collapse, they can affect whether the mass distribution *collapses to* its pseudo-Schwarzschild radius ("gravitational radius"; "Schwarzschild radius"). But they do not affect whether, once the mass distribution is entirely within its pseudo-Schwarzschild radius, a black hole forms. "The most general stationary ["stationary" means roughly: $t$ does not appear in the metric coefficients; less roughly: there is a timelike Killing vector] black hole solution to Einstein's equations is the analytically known Kerr-Newman metric. It is uniquely specified by just three parameters: the mass $M$, angular momentum $J$ and the [electric] charge $Q$ of the black hole. Special cases are the Kerr metric ($Q = 0$), the Reissner-Nordström metric ($J = 0$) and the Schwarzschild metric ($J = 0$, $Q = 0$)" (ref [5], p. 10). But it is not to be supposed that, for the purpose of assessing whether a mass distribution *becomes* a black hole, we are to replace the Schwarzschild exterior metric in an Oppenheimer-Snyder-like solution with some *other* metric; for example, in the case of a mass distribution with $Q = 0$ but $J \neq 0$, replacing it with a Kerr metric. It is to be emphasized that, even for a rotating mass distribution which becomes a Kerr black hole:

"The Kerr metric is not the exterior metric *during* the collapse; it is only the *asymptotic* form of the metric when all the dynamics has ceased" (ref [16], p. 359). Nor is the Reissner-Nordström metric the exterior metric during collapse to a Reissner-Nordström black hole. (It couldn't be, since it describes a *source-free* electric field; and the field can scarcely be source-free until after the black hole has formed.) In harmony with our principle, and the various remarks quoted above expressing it, we may say that, with respect to black-hole formation, the critical issue is still compaction within the pseudo-Schwarzschild radius; and charge and angular momentum enter the picture only with regard to *whether this compaction occurs* (and of course then also with regard to the properties of the black hole, if one forms).

For example, in the middle of p. 486 of ref [5], the displayed items are in the context of an assumed radial coordinate of *an* isotropic form in which M, that is, $GM/c^2$, is the Schwarzschild radius (cf. ref [16], p. 510; whereas with the Schwarzschild radial coordinate the Schwarzschild radius is of course $2GM/c^2$: cf. ref [16], p. 511). On p. 486 of [5] the issue being addressed is whether the centrifugal force due to the angular momentum of a collapsing rotating star will halt the collapse before a black hole (or any sort of singularity) forms. Where $R_b$ is the radius "at which the centrifugal force balances the gravitational force" the authors take for granted that if $R_b < M$ then "the star will already be inside a black hole before rotation can halt the collapse," and if $R_b > M$ then "the collapse will be halted at a radius larger than M, and no black hole forms." For our purposes, the important point here is that it is being taken for granted, even in this case of nonspherical collapse (i.e., absence of spherical symmetry; here because the star is rotating), that the critical radius of compaction with regard to black-hole formation is still the Schwarzschild radius. (This, even though, once a Kerr black hole has formed, the outermost Kerr radius is *smaller* than the Schwarzschild radius.)

I shall assume here that the pseudo-Schwarzschild radius (which is the "Schwarzschild radius" actually being addressed above) indeed plays this privileged role, even in cases of nonspherical collapse. There are very general analytic results perhaps suggestive of this, and there are results from numerical relativity. We begin with the analytic results. I quote again from Penrose (ref [15], pp. 104-106; for brevity I omit the important discussion on p. 106 of the fact that "the strict form of the reconverging light cone condition…that *every* ray through *p* should encounter sufficient material for divergence reversal to occur" is not really necessary "for the purposes of the singularity theorem being appealed to here"):

> Two familiar mathematical criteria for 'unstoppable collapse' are the existence of a *trapped surface* or of a point whose *future light cone begins to reconverge in every direction along the cone*. In either of these situations, in the presence of some other mild and physically reasonable assumptions, like the nonnegativity of energy (plus the sum of pressures), the nonexistence of closed timelike curves, and some condition of genericity (like the assumption that every causal geodesic contains at least one point at which the Riemann curvature is not lined up in a particular way with the geodesic), it follows (by results in Hawking and Penrose 1970 [ref [17]]) that a space-time singularity of some kind must occur. (Technically: the space-time manifold must be geodesically incomplete in some timelike direction.)…Basically, the argument for the physical realizability of the reconverging light cone condition is that given in Penrose (1969 [ref [18]]). Imagine a certain amount of massive material, say of total mass $M$, and allow it to fall to within a roughly defined region whose diameter is of the general order of $4GM$ [$= 4GM/c^2$; so radius $= 2GM/c^2$]. We consider a space-time point $p$ somewhere in the middle of this region, and examine the future light cone $C$ of $p$. Thus, $C$ is swept out by the future-endless rays (null geodesics) with past endpoint $p$. The strict condition that $C$ 'satisfies the reconverging light cone condition' would be that on every ray $\gamma$ generating $C$ there is a place where the divergence of the rays changes sign. If it is assumed that such a ray is geodesically complete in the future direction (and that the energy flux across the ray is nonnegative), then it follows that, to the future of $p$ along the ray, there is a point *conjugate* to $p$ (i.e. a point $q$, distinct from $p$, with the property that there is a 'neighbouring ray to $\gamma$' which intersects $\gamma$ in $p$ and again at $q$; more precisely, there is a nontrivial Jacobi field along $\gamma$ which vanishes both at $p$ and at $q$). The idea is that as the material falls in across $C$ it causes focussing of a sufficient degree that such divergence reversal indeed arises. This is merely a feature of there being enough 'focussing power' in the lensing effect of the Ricci tensor component along the ray (namely, $R_{ab}l^a l^b$, where $l^a$ is a null tangent vector to $\gamma$), due to the energy density in the matter falling in across $C$. There are simple integral expressions that can be written down (cf. Clarke 1993 [ref [19]], in particular) which provide sufficient conditions for a conjugate point to arise, so it is merely an order-of-magnitude requirement that there is sufficient infall of material to ensure that the focussing condition will be satisfied. The situation could be made to be qualitatively similar to the original Oppenheimer-Snyder (1939 [ref [3]]) collapsing dust cloud (pressureless fluid), but where there is no symmetry assumed and no particular equation of state employed (like that of Oppenheimer and Snyder's dust).

Though the phrases "of the general order of," "order-of-magnitude," and "qualitatively" make this merely harmonious with our widely-endorsed principle, it is striking that there is here no requirement of spherical symmetry (or of charge neutrality). And obviously the massive material need not *fall* to within the circumscribed region; there need only *be* "too much mass…concentrated in too small a volume." As we shall see, general relativity (assuming cosmic censorship, and assuming our principle) implies that in Euclidean space any Hubble volume which fails to be below average in mass density must be born a black hole. (There is an analogous result for hyperbolic space and *apparent Hubble volumes*; 'Hubble volume' and 'apparent Hubble volume' are precisely defined in the next section.) There will be no issue of *gravitational collapse* to within the pseudo-Schwarzschild radius; and thus no issues about charge or angular momentum. (Though I would in fact tend to assume, very plausibly given their nature, that for Hubble volumes [and apparent Hubble volumes] of the sizes we will be talking about, electric charge and angular momentum are negligible; that is, that [in square meters] $(GM/c^2)^2 >> (J/Mc)^2 + Q^2G/(4\pi\varepsilon_0 c^4)$, where M is mass in kg, J is angular momentum in $m^2$-kg/s, Q is electric charge in coulombs, and $1/4\pi\varepsilon_0$ is a constant [8.99 X $10^9$] in Newton-$m^2$[i.e. $m^3$-kg/$s^2$]/coulomb$^2$, $\varepsilon_0$ being the permittivity of free space.)

There is also Kip Thorne's *hoop conjecture* (ref [20]; ref [5], p. 353), which is tailored for the absence of spherical symmetry (and does not presuppose cosmic censorship): *Black holes with horizons form when and only when a mass M gets compacted into a region whose circumference in every direction is bounded by C ≤ 4πGM/c²*. (My statement of the conjecture may be controversial, since the inequality is generally taken merely as an approximation; but the evidence that exists for the conjecture is evidence for either version. The compacted mass $M_c$ is for technical reasons idealized as being the Arnowitt-Deser-Misner, or ADM, mass; see the discussion in Frolov and Novikov, [8], p. 190.) In the case of a sphere of radius *r'*, just sufficient to enclose a mass $M_c$, the hoop conjecture implies that $M_c$ is within a black hole if *r'* ≤ 2G$M_c$/c². So the hoop conjecture implies our principle. If the conjecture is restated, as it often is, because of worries about the teleological nature of the event horizon, in terms of a black hole's locally ascertainable *apparent horizon* (this is a *black hole* "apparent horizon"; it has nothing to do with the apparent horizon spoken of earlier), this presumably makes no difference here; for if there is an apparent horizon then it is within an event horizon (assuming the *weak energy condition* is satisfied; see ref [8], pp. 171, 178-179). There are results from numerical relativity which seem to support the hoop conjecture (see ref [5], pp. 353-357); and there are numerical results, for various forms of nonspherical collapse, which are just as good for our purpose. As Frolov and Novikov note (ref [8], p. 191), numerical simulations have obtained $C ≤ 15.8GM/c²$ and $C ≤ 16GM/c²$. Note that each is of the form $C ≤ α4πGM/c²$, where $α ≥ 1$ (in the first, $α = 3.95/π$; in the second, $α = 4/π$). These evidences support at least a *weak hoop conjecture*: *There is some α ≥ 1 such that: black holes with horizons form when and only when a* *mass M gets compacted into a region whose circumference in every direction is bounded by $C \leq \alpha 4\pi GM/c^2$.* For a Hubble volume/sphere of radius $R_{AH}$, $C = 2\pi R_{AH}$. Thus from the weak hoop conjecture there is some $\alpha \geq 1$ such that the Hubble volume is within a black hole if $R_{AH} \leq \alpha 2GM/c^2$. But we shall show below that (for any $\alpha \geq 1$) in Euclidean space the inequality is inevitable unless the mass density in the Hubble volume is below average.

I will assume here that either (i) the hoop conjecture, "or something very like it" (such as the weak hoop conjecture), is correct, or (ii) for some other reason (along the lines of what Penrose says) we do indeed want to maintain the (very widely supported) principle adopted above: If the current mass $M_c$ enclosed by a sphere of radius $r$' is such that $r$' $\leq r_{PS}$ then the region enclosed by the sphere is within a black hole. (For an entry into the literature, read ref [21]; then [22]; then the extremely difficult and fascinating [23]. In [21], p. 2, the author notes that "despite all difficulties [of formulation; which are many], the Hoop Conjecture has been very successful…Many numerical and/or analytical idealized examples…have given robust support to the conjecture.")

II. The Black Hole Next Door

We begin our argument with the supposition that the universe is now spatially flat. If the universe is now spatially flat, then the post-inflationary universe was always spatially flat. For it is well known that if at any earlier time in the post-inflationary history of the universe, it failed to be spatially flat, then it deviated thereafter more and more from flatness, and is not flat now. So, on our supposition that the universe is now spatially flat, we know that the post-inflationary universe was always flat.

At every time $t$ after, say, the epoch of decoupling (about 380,000 years after the Big Bang), at every point $(x, y, z)$ of space, the *Hubble volume* V$(x, y, z, t)$ is the sphere centered on $(x, y, z)$ whose radius is the *Hubble radius* $c/H_t$. If the post-inflationary universe is always spatially flat, then V$(x, y, z, t)$ extends to its apparent horizon, since (as we noted earlier) for a flat universe the apparent horizon of a Hubble volume coincides with the spherical boundary of the Hubble volume (that is, the radius $R_{AH}$ to the apparent horizon is the same as the Hubble radius).

By a *Hubble black hole* (HBH), I mean a Hubble volume which is within a black hole.

Consider our current Hubble volume, centered on (X,Y, Z) = the current center of mass of the Earth. This sphere has a radius of (where $H_0$ is the current value, i.e. the "Hubble constant," of the Hubble parameter) $c/H_0$, which is about $10^{26}$ m, or about 13.8 thousand million light-years. It is well known that, due to cosmological expansion, objects which appear to be, say, 13 thousand million light-years away are really about some 40 thousand million light-years away, and some authors would say that our current "Hubble volume" is thus some 40 thousand million light-years in radius. Note well that I am *not* using 'Hubble volume' that way. By "our Hubble volume" at $t$ I mean the sphere centered on (X,Y, Z) whose radius is $c/H_t$. Now, go back to a time $s$, 400,000 years after the Big Bang. Then "our Hubble volume" at $s$ is the sphere centered on (X,Y, Z) whose radius is $c/H_s$. Note that this is perfectly well defined, even though the Earth did not exist at $s$. (It is defined as being a certain sphere within our current Hubble volume.) What is the probability that, at $s$, there was adjacent to our Hubble volume at $s$, at least one HBH?

(Same-sized, of course; at any given time all Hubble volumes have the same size.) The probability is at least .99975. I will now explain why.

Consider an arbitrary Hubble volume $V(x, y, z, s)$ from that time, whose radius is $R_{AH}$. Since we are supposing that the universe is flat, there is no question that $(4/3)\pi R^3_{AH}$ is the proper volume of $V(x, y, z, s)$. Now, suppose, for the moment, that the mass density at $s$ in the Hubble volume is equal to the mass density $\rho_s$ of the universe at $s$. Then $\rho_s(4/3)\pi R^3_{AH}$ is the total mass $M_s$ enclosed in the Hubble volume. (Since we are interested only in the total mass in the volume, as a function of its total density, rather than as a function of its distribution, there is no need to integrate over expanding radii.) Thus

(4) $\qquad r_{PS} = [2G\rho_s(4/3)\pi R^3_{AH}]/c^2$

is the pseudo-Schwarzschild radius for $M_s$. For reasons already given, we have the fact that, if none of the mass $M_s$ enclosed by a sphere of radius $R_{AH}$ extends to a radius R' such that R' > $r_{PS}$, then the sphere is within a black hole. Thus, if $V(x, y, z, s)$ is not to be a HBH, it must be the case that:

(5) $\qquad R_{AH} > [2G\rho_s(4/3)\pi R^3_{AH}]/c^2$

or thus

(6) $\qquad R_{AH} > 8\pi G\rho_s R^3_{AH}/3c^2$

If the post-inflationary universe was always spatially flat, then we are obliged for well-known empirical reasons to construe the contribution, at any post-inflationary time $t$, from "dark energy" or the cosmological constant, as being part of $\rho_t$ (and in any event we are at liberty to do so; and there are indeed other reasons to take the view that the cosmological constant "describes a constant energy or mass per unit volume permeating the universe"; a contribution to "mass or energy density"; see ref [24], p. 449), and thus we use a standard form of the Friedmann equation in which no cosmological constant appears (it is "absorbed into" $\rho_t$):

(7)    $H_t^2 = (8\pi G/3)\rho_t - (kc^2/a_t^2)$

(There is a deduction of (7), from the Einstein field equations, in many places; for example, in ref [10], pp. 271-272.) By (7), with $k = 0$ and $t = s$, we have:

(8)    $\rho_s = 3H_s^2/8\pi G$

Thus, by (6) and (8):

(9)    $R_{AH} > H_s^2 R_{AH}^3/c^2$

But that is impossible, since, for $k = 0$ and $t = s$, $R_{AH} = c/H_s$. Thus, on our supposition that at $s$ the mass density in the region enclosed by the apparent horizon of $V(x, y, z, s)$ is equal to the mass density $\rho_s$ of the universe at $s$, $V(x, y, z, s)$ must be a HBH. In other words, Hubble volumes have a remarkable property: if the universe is spatially flat and the mass density in the Hubble volume is equal to the mass density of the universe, then the pseudo-Schwarzschild radius of the enclosed mass is simply the Hubble radius itself.

Furthermore, if at $s$ the mass density in the region enclosed by the apparent horizon of V($x, y, z, s$) is *greater* than $\rho_s$, then there is an additional factor $\alpha > 1$ in the numerators on the right sides of (4)-(6), and thus in the numerator on the right side of (9), and we again have an impossible inequality. The only way our arbitrary V($x, y, z, s$) can fail to be a HBH is if at $s$ the mass density in the region enclosed by its apparent horizon (that is, in the Hubble volume) is less than $\rho_s$; the Hubble volume must be underdense. We thus divide the Hubble volumes at time $s$ into those which are underdense (mass density $< \rho_s$) and those which are nonunderdense (mass density $\geq \rho_s$). The latter must be HBHs.

[Let me add a note concerning Sean Carroll's "The Universe Is Not a Black Hole" (online; just google in the title). Given any plausible form of inflation, at the time of decoupling the universe had long since expanded to a size at the very least as great as that of our current Hubble volume. So, with respect to Hubble volumes at $s$, there were plenty of places for a light ray not to be able to escape *to*. Unless either our principle about a condition sufficient for black-hole formation is incorrect, or space is not Euclidean, any nonunderdense Hubble volumes at $s$ were well and truly (within) black holes. And, with respect to our current Hubble volume (in of course our sense of 'Hubble volume'), it had better be underdense, if space is flat.

The proof of the "remarkable property" mentioned above is so elementary that I always wondered whether it had already been noticed. After writing the rest of this paper,

I chanced upon the piece by Carroll. (Please note that I am making no claim that the universe is a black hole! To what "exterior" region would the Schwarzschild metric be applied?) What Carroll says makes clear that he is aware, and that apparently others had been aware, of the property, and of the triviality of its proof. But, as we shall now see, it has non-trivial implications. The point is that the result holds at all times, as an intrinsic feature of general relativity. We may then consider ancient Hubble volumes, external to what was then our own Hubble volume, and consider whether they were HBHs. As we shall see, if space is Euclidean then (to an extreme probability) we *observe* unspeakably large black holes. We don't; so it isn't. (There is an analogous result for hyperbolic space; which holds to a probability high beyond human imagining.) This affords us an indirect observational means of determining the curvature of space, even if slow-roll inflation stretched it to superhorizon scales.]

Now, divide up the universe at $s$ (non-exhaustively, of course) into non-overlapping Hubble volumes, such that our own Hubble volume at $s$ is adjacent to twelve others (four to the sides, like a cross; four above in the low places, and four below; alternatively, place the centers of the twelve spheres "at the vertices of a regular icosahedron"; ref [25], p. 134). We assume that for an arbitrary Hubble volume at $s$ in this universe, it is 50/50 whether its mass density is below, or not below, the average mass density of the universe at $s$. Assume also, for the moment, independence. Then the probability that at least one of the twelve Hubble volumes is a HBH is $1 - (.5)^{12}$, which is slightly greater than .99975. But you may complain about our assumption of independence; perhaps whether one Hubble volume is underdense, or not, affects whether a neighbor is underdense, or not. However, given that our own Hubble volume was (of course) underdense, eleven of our neighbors also being underdense could only *enhance* the probability that the remaining neighbor is nonunderdense. So the probability that we had at least one HBH neighbor is *at least* .99975.

(The total universe is of course neither underdense [mass density < $\rho_s$] nor *overdense* [mass density > $\rho_s$]. We are assuming that the probability that a Hubble volume at *s* is nonunderdense, given simply that it is selected from the total universe, is .5. Consider any large region of the universe that is at least slightly overdense. The probability that a Hubble volume is nonunderdense, given that it is selected from *that* region, is greater than .5. Just as, if the probability that you hit a red area, given simply that you hit the target, is .5, then the probability that you hit a red area, given that you hit the reduced target—from which a bit of whiteness is excluded, so that it is slightly "overred"—is greater than .5. Now, just focus on any two adjacent non-overlapping Hubble volumes, *A* and *B*. If *A* is underdense, then the universe outside of *A* must be at least slightly overdense. Thus, given that *A* is underdense, *B* is perforce selected from an at least slightly overdense remaining universe. Thus the probability that *B* is nonunderdense, given that *A* is underdense, is greater than .5. *A*'s being underdense enhances the probability of *B*'s being nonunderdense.)

Hubble volumes grow and merge with all or parts of their erstwhile neighbors into larger Hubble volumes, then having new, equally larger, neighbors. Now, select five times from the remote past, where each successor is greater than its predecessor by a factor of 4, more than allowing for our expanded Hubble volume of the later time to have absorbed all of its erstwhile immediate neighbors from the earlier time. Thus:

(i)  400,000 years after the Big Bang

(ii)  1.6 million years after the Big Bang

(iii) 6.4 million years after the Big Bang

(iv) 25.6 million years after the Big Bang

(v)  102.4 million years after the Big Bang

Our argument above applies at all five times: at each of these times there will be a probability of about .99975 that we acquire at least one HBH neighbor. So the probability that we acquire at least one HBH neighbor, at at least one of those five times, will be (since at each of the five times the probability of failing to have any HBH neighbor at that time is only .00025) $1 - (.00025)^5 \approx .999999999999999999$; or virtual certainty.

Almost certainly, then (if the universe is now spatially flat, and thus the post-inflationary universe was always spatially flat), we at some stage in the enlargement of our own Hubble volume (miraculously always underdense) acquired a less fortunate, and at that time equally large, HBH neighbor. Once a Hubble volume is a HBH, *that region* is *always* (on the timescales we are talking about) within a black hole, even when absorbed into a new, larger Hubble volume that manages to avoid being a HBH. And that enormous region, that is within a black hole, remains about as close to us as it always was, even when our own new, much larger Hubble volume of today includes it. (The distances we are talking about are small enough fractions of the current Hubble radius that we may ignore cosmological expansion. It wouldn't make much difference anyway.) In other words, if the universe is now spatially flat then it seems an essential certainty that there is in our sky a black hole of unspeakable size, and in astronomical terms not all that far away. Since, *x* years after the Big Bang the diameter of a Hubble volume is 2*x* light-years, the HBH neighbor is such that either:

(i)   it is about 400,000 light-years away, with a diameter of about 800,000 light-years, or

(ii)  it is about 1.6 million light-years away, with a diameter of about 3.2 million light-years, or

(iii) it is about 6.4 million light-years away, with a diameter of about 12.8 million light-years, or

(iv)  it is about 25.6 million light-years away, with a diameter of about 51.2 million light-years, or

(v)   it is about 102.4 million light-years away, with a diameter of about 204.8 million light-years

The pattern here is of course: it has a diameter of *D*, and is a distance away of *D*/2. In flat space, the angle θ, in degrees, subtended in the sky by a sphere of radius *r*, where the distance from us to the center of the sphere is *d*, is 2 arcsin (*r*/*d*). (Think of looking at the sphere as looking at a circle edge-on in a plane bisecting the sphere; as an inhabitant of Flatland might see it. In your mind, draw a tangent line segment EC from your eye to the circle. Draw the line segment CO from C to the center O of the circle = the center of the sphere. By Euclid's proposition III.16, < ECO is a right angle. In your mind, flip all of this over. Then you have a tangent line segment EK from your eye to a point K on the other side of the circle, and a line segment KO from K to the center of the circle. < EKO is also a right angle. Thus we have two mirror-image right triangles whose shared hypotenuse EO is the distance $d$ from your eye to the center of the circle = the center of the sphere, where CO = KO = the radius of the circle = the radius $r$ of the sphere. So the angle subtended by the circle-seen-edge-on, which is the same as the angle subtended by the sphere, is 2 arcsin ($r/d$).) Here $r = D/2$ and $d = D$. So our HBH subtends in the sky an angle of 2 arcsin (.5) = 60 degrees; looking up at the right time an enormous backdrop of total blackness behind a third of the sky, and no more than about 102.4 million light-years away. Such is not observed.

(What *would* be observed? Deflection of light rays by the black hole might "obscure," so to speak, part of the zone of darkness. See the startling "photograph" of a black hole, in ref [26], p. 31; that is, the computer-generated "simulation of the photographic appearance of a black hole surrounded by a disk of luminous gas." One sees about two-thirds of the black hole itself, the bottom part being cut off by light from the luminous gas, being bent around it. The whole thing looks rather like a mushroom, with about two-thirds of a disc of perfect blackness at its center. Because of the rotation of the luminous gas, and the optical deformation caused by the black hole, there is a pronounced left-right asymmetry in the light flux, and other peculiar properties. Of course, this is for a black hole of much more modest size than we are here envisaging. But, with respect to our argument, what we might lose in terms of the "obscuring by light" of part of an observationally unknown enormous backdrop of darkness, we gain in terms of an observationally unknown enormous *deformation of light*. No such enormous mushroom-caps of deformed light are seen in the sky. So for simplicity I will go on speaking as though the whole region of darkness would be seen as a backdrop.)

If the universe is now spatially flat then we see a distant black curtain of unspeakable size. We don't; so it isn't. The only way out would be to say that our location in the universe is special. Not only is the region of our current Hubble volume underdense; it is rather uniformly underdense, so that at no stage in the process of Hubble-volume expansion did the Hubble volume of that stage fail to be underdense, thus trapping us in a black hole. (So that, if the universe is now flat, we have been very lucky.) But it was also *very* uniformly underdense, so that at no point did we acquire a HBH neighbor.

One *could* say this: that we are living in a humongous density perturbation, and the universe is after all spatially flat. If so, our descendants must some day see in the sky a zone of growing darkness.

Our universe thus could be spatially flat, but surely we must nonetheless say that this is extremely improbable. Let $F$ = "The universe is now spatially flat" and $N$ = "No enormous backdrop of total blackness is seen in the sky." The *antecedent* conditional probability, $\Pr(F/N)$, is by Bayes's theorem:

(10)     $\Pr(F/N) = [\Pr(F) \text{ X } \Pr(N/F)]/\Pr(N)$

where $\Pr(F)$ is the prior probability of $F$ (which I shall assume is 1/3; that would seem to be the standard prior subjective probability), $\Pr(N)$ is the prior probability of $N$ (which I shall assume is 1/2; that is the minimum, since either one's prior is "50/50" or it reflects the view that there are more ways of there *not* being, than of there being, the span of uniformity associated with an enormous backdrop of total blackness), and $\Pr(N/F)$ is the antecedent conditional probability of $N$ given $F$, which by our argument is about $10^{-18}$. So the (antecedent) conditional probability of $F$ given $N$ is:

(11)     $\mathrm{Pr}(F/N) = [(1/3) \times 10^{-18}]/(1/2) \approx .667 \times 10^{-18}$

So when we learn, as we have learned, that $N$ is true, our updated $\mathrm{Pr}(F)$ should be about .667 X $10^{-18}$. That is, we should say that there is less than one chance in a million trillion that space is Euclidean.

Though I will not give the argument in full, we can get the same extreme improbability that space is *hyperbolic* (negatively curved; $k < 0$). By an *apparent Hubble volume* $\mathrm{AV}(x, y, z, t)$ we mean the sphere centered on $(x, y, z)$ whose radius is $c/[\mathrm{H}_t^2 + (kc^2/a_t^2)]^{1/2}$; we shall just call it $\mathrm{V}(t, \mathrm{R_{AH}})$. (In flat space, these coincide with Hubble volumes; in hyperbolic space apparent Hubble volumes are bigger than the corresponding Hubble volume; in spherical space apparent Hubble volumes are smaller.) By an *apparent Hubble black hole* (AHBH) we mean an apparent Hubble volume that is within a black hole. For reasons already given, if an arbitrary apparent Hubble volume $\mathrm{V}(t, \mathrm{R_{AH}})$ is not to be an AHBH, then $\mathrm{R_{AH}}$ must be greater than the pseudo-Schwarzschild radius of the mass $\mathrm{M_c}$ it encloses. Suppose, at time $t$, that the mass density of the apparent Hubble volume is equal to the mass density $\rho_t$ of the universe at $t$. Then, if $\mathrm{V}(t, \mathrm{R_{AH}})$ is not to be an AHBH, we must have:

(12)     $\mathrm{R_{AH}} > 2\mathrm{G}\rho_t V/c^2$

where $V$ is the proper volume of V($t$, R$_{\text{AH}}$). Where the line element in comoving coordinates ($t$, $r$, θ, φ) of a FLRW spacetime is given by:

(13)     $ds^2 = -c^2dt^2 + a_{\text{t}}^2[dr^2/(1 - kr^2) + r^2(d\theta^2 + \sin^2\theta d\varphi^2)]$

and where R $=_{\text{df}} a_{\text{t}} r$, the proper volume of a sphere of radius R (whether space is flat, hyperbolic, or spherical) is given by (ref [6], p. 3):

(14)     $V = \int_0^{2\pi} d\varphi \int_0^{\pi} d\theta \int_0^r dr' \, [g^{(3)}]^{1/2} = 4\pi a_{\text{t}}^3 \int_0^{\chi} d\chi' \, f^2(\chi')$

($r'$ and χ' are integration variables) "where $g^{(3)} = (a_{\text{t}}^6 r'^4 \sin^2\theta)/(1 - kr'^2)$ is the determinant of the restriction of the metric $g_{\text{ab}}$ to the 3-surfaces $r'$ = constant," and "where χ is the hyperspherical radius": that is, (13) can be rewritten as $ds^2 = -c^2d\tau^2 + a_\tau^2[d\chi^2 + f^2(\chi)(d\theta^2 + \sin^2\theta d\varphi^2)]$ "where τ is proper time on comoving world lines (along which χ, θ, and φ are all constant)" (ref [4], p. 91; cf. ref [27], p. 255) such that $r = f(\chi) =$ (i) sinh χ if $k < 0$, (ii) χ if $k = 0$, (iii) sin χ if $k > 0$ (ref [6], p. 3; ref [27], pp. 256-258). If $k = 0$, $V$ turns out to be (4/3)πR$^3$ (ref [6], p. 3). Since (for χ ≠ 0) $\sinh^2 \chi > \chi^2 > \sin^2 \chi$, we see from (14) that, when $k < 0$:

(15)     $V > (4/3)\pi\text{R}^3$

(In a spatially hyperbolic FLRW spacetime a sphere of radius R has a greater volume than in a spatially flat FLRW spacetime; and, we note, greater in a spatially flat FLRW spacetime than in a spatially spherical one. Our arguments cannot impact spherical spaces. They turn crucially on the fact that in Euclidean and hyperbolic spaces the volume of the sphere is $\geq (4/3)\pi R^3$; which is too great.)

By (12) and (15):

(16)     $R_{AH} > [2G\rho_t(4/3)\pi R^3_{AH}]/c^2$

By the Friedmann equation (7) we have:

(17)     $\rho_t = (3/8\pi G)(H_t^2 + kc^2/a_t^2)$

Thus by (16) and (17):

(18)     $R_{AH} > (H_t^2 + kc^2/a_t^2)R^3_{AH}/c^2$

But that is impossible, since at $t$ $R_{AH} = c/(H_t^2 + kc^2/a_t^2)^{1/2}$. Thus, on our supposition that at $t$ the mass density of the apparent Hubble volume $V(t, R_{AH})$ is equal to the mass density $\rho_t$ of the universe at $t$, $V(t, R_{AH})$ must be an AHBH.

Furthermore (as before), if at $t$ the mass density in the apparent Hubble volume is *greater* than $\rho_t$, then there is an additional factor $\alpha > 1$ in the numerators on the right sides of (12), (16), and thus (18), and we again have an impossible inequality. The only way our arbitrary apparent Hubble volume $V(t, R_{AH})$ can fail to be an AHBH is if at $t$ its mass density is less than $\rho_t$; the apparent Hubble volume must be underdense.

We could then argue, in a way very similar to what we did before, from the supposition that space is now hyperbolic to an extreme probability that there is an enormous backdrop of total blackness in the sky. For, if space is now hyperbolic, then the post-inflationary universe must always have been spatially hyperbolic. For if the post-inflationary universe was ever (precisely) flat, then it is flat now, and not hyperbolic. And if the post-inflationary universe was ever spatially spherical then it deviated more and more *away* from flatness (became more and more spherical), and so is not hyperbolic now; the only path from positive curvature to negative curvature perforce moving through flatness. Thus, if space is now hyperbolic, the post-inflationary universe was always hyperbolic.

Since, if space is hyperbolic, the current negative curvature $k/a_t^2$ (in inverse length-squared, $= kc^2/a_t^2$ in inverse time-squared) is constrained by observation to be extremely small (and would have to have been smaller still at earlier post-inflationary times, since a spatially hyperbolic universe tends to greater negative curvature over time), an apparent Hubble volume (radius $c/(H_t^2 + kc^2/a_t^2)^{1/2}$) is only very slightly bigger than the corresponding Hubble volume (radius $c/H_t$). Thus, remembering that our factor of 4 was generous, *more* than allowing for our expanded Hubble volume of the later time to have absorbed all of its erstwhile adjacent Hubble neighbors from the earlier time, we may then use the same five times in the remote past as we did before. (I shall not here go into the subtleties about kissing spheres in hyperbolic space; I will just note that the hyperbolic space is at each time partially divisible into non-overlapping apparent Hubble volumes in such a way that our own apparent Hubble volume had twelve apparent-Hubble-volume neighbors.) At each time, our own apparent Hubble volume has at least a

.99975 probability of acquiring an AHBH neighbor. Thus, as before, the probability will be .999999999999999999 that we acquired an AHBH neighbor at at least one of those five times, and thus virtually certain that there is now an enormous zone of darkness in the sky. Thus, where $N$ ("No enormous backdrop of total blackness is seen in the sky") and $F$ ("The universe is now spatially flat") are as before, and $H$ = "The universe is now spatially hyperbolic," the antecedent $\Pr(N/H)$, like the antecedent $\Pr(N/F)$, = $10^{-18}$.

We now give a new Bayesian argument. Let $N$, $F$, $H$, be as above, and let $S$ = "The universe is now spatially spherical." For a FLRW universe, $F$, $H$, and $S$ are exclusive and exhaustive. Thus $\Pr(S) = 1 - \Pr(F \text{ v } H)$. We assume that (the prior) $\Pr(F)$ = $\Pr(H) = \Pr(S) = 1/3$, and that (the prior) $\Pr(N) = 1/2$. We have argued that (the antecedent) $\Pr(N/F) = \Pr(N/H) = 10^{-18}$. Thus the antecedent conditional probability of $N$ given $F$ v $H$ is:

(19)     $\Pr(N/F \text{ v } H) = \Pr(N/F) = \Pr(N/H) = 10^{-18}$

(It is a theorem of the probability calculus that if $\Pr(A/B) = \Pr(A/C)$ and $\Pr(B) = \Pr(C)$ and $\Pr(B \ \& \ C) = 0$ then $\Pr(A/B \text{ v } C) = \Pr(A/B) = \Pr(A/C)$.) Thus the antecedent conditional probability of $F$ v $H$ given $N$ is by Bayes's theorem:

(20)     $\Pr(F \text{ v } H/N) = [\Pr(F \text{ v } H) \ X \ \Pr(N/F \text{ v } H)]/\Pr(N)$

         $= [(2/3) \ X \ 10^{-18}]/(1/2) \approx 1.334 \ X \ 10^{-18}$

So when we learn, as we have learned, that $N$ is true, our updated probability for $F$ v $H$ should be about $1.334 \times 10^{-18}$. Thus the probability that the universe is now spatially spherical (and thus spatially finite) is about .999999999999999998666.

We could in fact tweak the above arguments to get still higher probabilities. We could select up to *eight* times in the remote past (ending with 6, 553.6 million years after the Big Bang), since taking cosmological expansion into account could scarcely take the sting out of the enormity of the zone of darkness that would be seen in the sky. It could reduce it only to something like .5 radians; but even .5 radians is still over 28 degrees of the celestial circle. (With regard to $N$—"No enormous backdrop of total blackness is seen in the sky"—let us make the modest assumption, which would be readily accepted by astronomers, that anything greater than 10 degrees counts as "enormous." Note that the Moon, even when closest to the Earth, subtends an angle of only a little more than *half* a degree [≈ .5548 degrees], less than a *hundredth* of a radian. Our black hole would have an angular size more than 50 times that of the Moon.) If we go to the eight times in the past, then the antecedent $\Pr(N/F) = \Pr(N/H) = \Pr(N/F$ v $H)$ becomes about $1.526 \times 10^{-29}$, the antecedent $\Pr(F$ v $H/N)$ becomes by Bayes's theorem about $2.035 \times 10^{-29}$, and thus the updated $\Pr(F$ v $H)$ becomes about $2.035 \times 10^{-29}$ and the probability that the space around you is positively curved is then about .99999999999999999999999999997965.

But this is understatement. Let us again consider the hypothesis that the spatial sections of our spacetime have a hyperbolic geometry. Let $Z$ = "There is no disc-shaped zone of total blackness (through which absolutely nothing can be seen) beginning at a radial distance of no more than about $10.35 \times 10^9$ light-years away, and subtending in our sky an angle of at least about 38.9 degrees (so that the area blacked out, about 1187 square degrees, is about twice the size of the Orion constellation)." Without assuming that we have learned anything more controversial than that *Z* is true, the hypothesis that space is hyperbolic can be ruled out; the probability that space is not hyperbolic becomes unspeakably high.

Go back to a time *u*, 3,450 million years after the Big Bang. Call the *apparent Hubble volumes* from time *u*, *Ubble volumes*. In hyperbolic space, how many non-overlapping Ubble volumes will *our current Hubble* volume have absorbed? Recalling that negative curvature is constrained by observation to be extremely small at all post-inflationary times, the difference between the radius of an Ubble volume and that of the corresponding Hubble volume from time *u* will be negligible: the radius of the Ubble volume will be only very slightly greater than 3,450 million light-years; and we will just call it about 3,450 million light-years. For same-sized spheres the *maximum* sphere-packing density in hyperbolic *space* (as opposed to sphere-packing in, say, a larger *sphere*) is about 85% (ref [28]); that is, at most about "85% of the space" can also be within the packed spheres. (The maximum for Euclidean space is, by Kepler's conjecture [ref [29]], for which the computer-assisted proof by Hales [refs [30]–[37]] is widely accepted, $\pi/(18)^{1/2}$, or about 74%.) The maximum density in hyperbolic space for sphere-packing *in a larger sphere* may well be not only less than 85% but less than even $\pi/(18)^{1/2}$ (refs [38], p. 9 [the "Strong Kepler Conjecture"]; [39], pp. 12-13). In any event, these are *maxima*; and the subtleties concerning "edge effects" for packings within bounded containers are notorious. (Stewart, ref [40], p. 75, says that "for instance, hardly anything worthwhile is known about the most efficient way of packing spheres into ordinary box-shaped boxes.") We shall here, with what may seem a strange combination of over-caution and indifference, take just a *tenth* of the volume of our current Hubble volume in hyperbolic space as being available to be exhaustively divisible into non-overlapping Ubble volumes. (As we shall see, *it really doesn't matter*. It's probably much greater than a tenth; but feel free to take a thousandth, or a trillionth, or a googolth.)

Where $K_t$ is the absolute value of the negative curvature $k/a_t^2$ of the hyperbolic space at time *t* (that is, of the constant-time spatial section of the spacetime, where the time is *t*), we may express the volume at *t* of a sphere of radius R as (ref [41], p. 7; the formula appears in the great [42]):

(21)    $[\pi K_t^3 \sinh (2R/K_t)] - 2\pi K_t^2 R$

(Since $\sinh(-x) = -\sinh(x)$, obviously (21) has the same value whether we take $K_t$ to be $k/a_t^2$ or its absolute value. It is convenient to do the latter.) Measuring radii in light-years, in hyperbolic space the available amount (exhaustively divisible into Ubble volumes) of the volume of our current Hubble volume of radius about $1.38 \times 10^{10}$ is thus we are assuming (where *t* = 0 is today and K = $K_0$):

(22)    $(.1) \{[\pi K^3 \sinh (2.76 \times 10^{10}/K)] - 2\pi K^2 (1.38 \times 10^{10})\}$

And the volume of an Ubble volume of radius about $3.45 \times 10^9$ is (in today's hyperbolic space):

(23)    $[\pi K^3 \sinh (6.9 \times 10^9/K)] - 2\pi K^2 (3.45 \times 10^9)$

We are interested in how many times (23) goes into (22). First divide both by $\pi K^3$. We then want to divide

(24)    (.1) $\{[\sinh (2.76 \text{ X } 10^{10}/\text{K})] - (2.76 \text{ X } 10^{10}/\text{K})\}$

by

(25)    $[\sinh (6.9 \text{ X } 10^{9}/\text{K})] - (6.9 \text{ X } 10^{9}/\text{K})$

We note that, since K is a very small number, the arguments of the hyperbolic sine function in (24) and (25) are both extremely large. The value of $\sinh (x) =_{\text{df}} (e^x - e^{-x})/2$ increases exponentially for increasing $x$. So for very large $x$ the difference between $\sinh(x)$ and $\sinh(x) - x$ is negligible. So to a close approximation we may divide

(26)    (.1) $\sinh (2.76 \text{ X } 10^{10}/\text{K})$

by

(27)    $\sinh (6.9 \text{ X } 10^{9}/\text{K})$

For very large $x$, $\sinh(x) \approx e^x/2$. We may thus divide

(28)    (.1) $e^{27600000000/\text{K}}/2$

by

(29)     $e^{6900000000/\text{K}}/2$

which is (.1) $e^{20700000000/\text{K}} > $ (.1) $e^{20700000000} > $ (.1) $2^{20700000000}$. We choose the last as our minimum, which is in the ballpark of (.1) $10^{6210000000} = 10^{6209999999}$. In hyperbolic space, this is the minimum number of non-overlapping Ubble volumes in our current Hubble volume. Our own ancient Ubble volume will have expanded so that our current Hubble volume has absorbed (about; one less than) $10^{6209999999}$ external Ubble volumes. That is, there are at least about $10^{6209999999}$ Ubble volumes in our sky. (This is of course more than would be the case in Euclidean space. The reason for the stupendous difference is that a sphere of radius 1.38 X $10^{10}$ light-years has a *vastly* greater volume in hyperbolic space— "Big Sky" country—than in Euclidean space, swamping the also greatly increased volumes of about-3450-million-light-year-radii Ubble volumes in hyperbolic space.)

In a hyperbolic space of very small negative curvature, an Ubble volume has a diameter of about 6,900 million light-years. And it can be no more than about 6,900 million light-years away. (Remember that we are talking about Ubble volumes within our current 13,800-million-light-year-radius Hubble volume.) I shall assume here that the negative curvature is so small that we may approximate the angular size of a sphere of radius $r$, where $d$ is the distance to the center of the sphere, as its value in Euclidean space: 2 arcsin $(r/d)$. Here $r = 3.45$ X $10^9$ and $d = 10.35$ X $10^9$, so we may approximate the angular size of the Ubble volume as 2 arcsin (3.45 X $10^9$/10.35 X $10^9$) = 2 arcsin (1/3) ≈ 38.9 degrees. (Such an Ubble volume, if also an AHBH, would render $Z$ false: there would be a disc-shaped zone of total blackness [through which absolutely nothing can be seen] beginning at a radial distance of no more than about $10.35 \times 10^9$ light-years away, and subtending in our sky an angle of at least about 38.9 degrees [so that the area blacked out, about 1187 square degrees, is about twice the size of the Orion constellation].)

Assuming independence (and that space is hyperbolic; and that for each of the external Ubble volumes it is 50/50 whether its mass density at $u$ was below, or not below, the average mass density of the universe at $u$), the probability that all of these $10^{6209999999}$ Ubble volumes were underdense at $u$ is .5 to the power of $10^{6209999999}$, which is in the ballpark of 10 to the power of $-(3 \times 10^{6209999998})$. And thus (recalling that in hyperbolic space the only way an apparent Hubble volume can fail to be an AHBH is by being underdense) $\Pr(Z/H) = 10$ to the power of $-(3 \times 10^{6209999998})$. Thus, assuming that the prior $\Pr(H) = 1/3$ and that the prior $\Pr(Z) = 1/2$ (the latter is again a minimum, since either one's prior is "50/50" or it reflects the view that there are more ways of there *not* being, than of there being, the span of uniformity associated with an enormous backdrop of total blackness), by Bayes's theorem the antecedent $\Pr(H/Z) = [(1/3) \times 10$ to the power of $-(3 \times 10^{6209999998})]/(1/2) \approx 10$ to the power of $-(3 \times 10^{6209999998})$. Thus, when we learn, as we have learned, that $Z$ is true, our updated probability for $H$ should be 10 to the power of $-(3 \times 10^{6209999998}) \approx$ one chance in googolplex to the power of $10^{6209999898} =$ ain't no way.

In short, the probability that space is hyperbolic is not interestingly distinguishable from zero. This has important implications. Let me just quote a remark by Leonard Susskind (ref [43], p. 371; emphasis in original): "…bubble nucleation has a distinct signature. If our pocket universe was born in a bubble-nucleation event, the universe must be *negatively curved*." (Susskind is alluding to ref [44], p. 3311.)

[Let me mention, in passing, a simpler argument against space being hyperbolic; one not dependent on any consideration of black holes, and decisive all by itself. Suppose that space is hyperbolic, to some small degree, compatible with the well-known evidence. Then an Ubble volume has a radius of about $3.45 \times 10^9$ light-years, or roughly $10^{25}$ m. We know from (14) above that in hyperbolic space the volume of a sphere of radius R is greater than $(4/3)\pi R^3$. Thus, in hyperbolic space, the volume of an Ubble volume is greater than $10^{75}$ m$^3$. Assume now, what is so cautious as to be beyond any conceivable dispute, that at least a googolth of the volume of our current Hubble volume can be exhaustively divided into non-overlapping Ubble volumes. Then we see, by an argument like that above, that there are at least $10^{6209999900}$ *non-overlapping* Ubble volumes in our current Hubble volume. Thus, in hyperbolic space, the volume $V_H$ of our current Hubble volume is at least $10^{6209999975}$ m$^3$. Ignoring cosmological expansion (which would make little difference), we know from WMAP and other observations that the mass density $\rho_H$ in our current Hubble volume is *very nearly* $3H_0^2/8\pi G$. Thus, where $M_H$ is the mass in our current Hubble volume, $M_H/V_H \approx 3H_0^2/8\pi G$. $H_0$ is supposed to be roughly $73[(km/s)/Mpc]$. A megaparsec (Mpc) is about $3 \times 10^{22}$ m. So $H_0 \approx (73 \times 10^3$ m/s)/(3 $\times$ $10^{22}$ m) $\approx 24.3 \times 10^{-19}$ s$^{-1}$. (We see, by the way, that H can be measured in units of s$^{-1}$, which is why $c$/H is a length.) G is about $6.673 \times 10^{-11}$ m$^3$/s$^2$-kg. Thus $3H_0^2/8\pi G \approx 10^{-26}$ kg/m$^3$. Thus $M_H/10^{6209999975}$ m$^3 \approx 10^{-26}$ kg/m$^3$. Thus, if space is hyperbolic, the mass $M_H$ in our current Hubble volume $\approx 10^{6209999949}$ kg! It is transparently obvious that the unspeakably vast proportion of this mass must be in the form of (positive) vacuum energy. Thus the proportion of vacuum energy density to matter density is unspeakably large. Assuming that the situation in our region of hyperbolic space is not *unspeakably* *unrepresentative* of that elsewhere in hyperbolic space (where, elsewhere, space being hyperbolic should have a similar effect), vacuum energy density in the universe is unspeakably greater than matter density. Not *half* as great, as in Einstein's model of a static (albeit, as it turns out, unstable) universe; not *three times* greater, as in the standard cosmological model of a universe thus now undergoing accelerated expansion; but *unspeakably many times* greater. It is then understatement to say that the repulsive gravity due to the density of positive vacuum energy overwhelms the gravitational attraction due to the density of matter. And so the universe rips itself apart.]

Now let us turn again to the hypothesis that the spatial sections of our spacetime are Euclidean. Let $M$ = "At no radial distance between here and the last-scattering surface is there a slightly-more-than-Moon-sized (about 37 arcminutes; the Moon's maximum angular size is about 33 arcminutes) zone of total blackness through which absolutely nothing can be seen." Though $M$ is not remotely so obvious as $Z$, I shall here assume that we have learned, or will learn, that $M$ is true. A negative answer to the following question has, as we shall now see, important implications, but I just don't know the answer, and would appreciate it if someone would tell me: Is there a Moon-sized curtain of complete blackness somewhere in the sky? Note that we are talking about something much more startling than what astronomers call a *void*. Voids simply "contain a much lower than average density of galaxies" (ref [13], p. 320); they do not obscure the heavens beyond. One naturally supposes that a distant 37-arcminute zone of total blackness would block our view of part of the last-scattering surface in some way contrary to experience. But the subtleties about foregrounds, and their subtraction from, or affect on, the cosmic microwave background (CMB), are complex (see ref [45]). Nevertheless, the Planck satellite has an angular resolution of 5 arcminutes (improved over the 15 arcminute resolution of the Wilkinson Microwave Anisotropy Probe). Setting aside issues about boundary effects concerning our distant 37-arcminute black-Moon curtain, focus on a 15-arcminute region *in its middle*. When foreground microwave emitters (between us and the curtain) are subtracted, Planck would either see some sort of 15-arcminute "gap" in the CMB, or register an enormous temperature fluctuation, since a black hole of the size we are talking about would have a temperature barely above absolute zero (stunningly less than the average 2.7 Kelvin of the CMB; the anomalous 300-arcminute Cold Spot in the CMB, confirmed by Planck, would, temperature-wise, pale in comparison). I shall assume here that, for a HBH between us and the last-scattering surface, if the HBH subtends in our sky an angle of (at least) 37 arcminutes then its distorting effects on the CMB would have been noticed by Planck.

Go back to a time $v$, 75 million years after the Big Bang. Call the Hubble volumes of that time *Vubble volumes*. In Euclidean space, how many non-overlapping Vubble volumes will our own expanding Hubble volume have absorbed between then and now? We can get a conservative minimum in the following way. Our current Hubble volume is a sphere. Inscribe in it a cube. Since, measuring in light-years, the diameter of the sphere is about $2.76 \times 10^{10}$, the edges of our cube are $2.76 \times 10^{10}/(3)^{1/2} \approx 1.59 \times 10^{10}$. A little cube of edge-length $1.5 \times 10^{8}$ is a box just big enough to contain a Vubble volume. Since $1.59 \times 10^{10}/1.5 \times 10^{8} = 106$, the cube inscribed in the sphere will hold $106^{3} = 1{,}190{,}016$ of the little cubes, and thus our current Hubble volume will contain at least $1{,}190{,}016$ non-overlapping Vubble volumes. Thus our own ancient Vubble volume will have expanded to absorb at least 1,190,015 external Vubble volumes. That is, there are at least 1,190,015 Vubble volumes in our sky.

A Vubble volume has a diameter of 150 million light-years, and obviously it can be no more than about 13,800 million light-years away. (Remember that we are talking about Vubble volumes within our current Hubble volume.) Calculating from a (roughly) "most-distant-case" scenario, and noting that the distance to the Vubble volume is then much greater than its diameter, the angle subtended in the sky is in radians approximately the diameter divided by the distance, and so is at least .010869 radians ≈ .6226 degrees (= 37.356 arcminutes), which is more than Moon-sized (≈ .5548 degrees at perigee). Assuming independence (and that space is Euclidean; and that for each of the external Vubble volumes it is 50/50 whether its mass density at $v$ was below, or not below, the average mass density of the universe at $v$), the probability that all of the (at least) 1,190,015 external Vubble volumes in our sky were underdense at $v$ is $(.5)^{1190015}$ which is in the ballpark of $10^{-357000}$. And thus (recalling that in flat space the only way a Hubble volume can fail to be a HBH is by being underdense) $\Pr(M/F) = 10^{-357000}$. Thus, assuming that the prior $\Pr(F) = 1/3$ and that the prior $\Pr(M) = 1/2$ , by Bayes's theorem the antecedent $\Pr(F/M) = [(1/3) \text{ X } 10^{-357000}]/(1/2) \approx 10^{-357000}$. Thus when we learn, as I assume we have learned, or will learn, that $M$ is true, our updated probability for $F$ should be about $10^{-357000}$ and the probability that space is non-Euclidean becomes a decimal point followed by three hundred fifty-seven thousand 9s.

The probability that, in a universe approximated on large scales as a FLRW spacetime, such as ours is supposed to be, space has negative curvature, is essentially zero; let's just call it zero. So the probability that space has positive curvature is the same as the probability that it is non-Euclidean; that is, Pr(*S*) = 1 − Pr(*F*), Pr(*H*) now being zero. So, assuming that we have learned or will learn that *M* is true, the probability that space is positively curved (and thus finite) is something like a decimal point followed by three hundred fifty-seven thousand 9s. Obviously space is not hyperbolic. If we learn that *M* is true then we know that space is spherical. If we do not learn that *M* is true, then we have one chance in a hundred thousand trillion trillion that space is not spherical.

The angles in a little triangle on your desk add up to more than π. Otherwise there would be monsters in the sky.

III. The Unreality of the Things around You

The reality of the things around you can scarcely depend on whether the universe happens to be spatially flat. So I shall give my argument here (in its initial formulations) on the pretence that it is (it makes the mathematics of volumes much simpler).

To do what we want to do, we do not need "the holographic principle" in general, nor even the "spherical entropy bound" for restricted regions of space, such as spheres smaller than the apparent horizon (see refs [46]-[48]). All we need is the holographic principle, or the spherical entropy bound, *for black holes*. By the Bekenstein-Hawking formula the entropy of a black hole is:

(30)     $S_{BH} = c^3 k_B A / [4G(h/2\pi)] = A/4$

where *A* is the area of the event horizon in Planck units (and where $c = G = h/2\pi = k_B = 1$). (Ref [12], p. 563: "This relation, known as the *Bekenstein-Hawking (BH) entropy* *formula*, appears to be universally valid (for any black hole in any dimension), at least when $A$ is sufficiently large." For "small black holes," where [p. 603] "terms of higher-order than the Einstein-Hilbert term contribute to the action in a significant way, the BH entropy formula is no longer correct"; and one turns to a generalization, the Wald entropy formula. See refs [49] –[51]. This does not affect our argument below, since, for the black holes we discuss, the horizon area is quite sufficiently large! For "large black holes," Wald entropy reduces to Bekenstein-Hawking entropy. *Even if it didn't*, it would make no difference to our argument, since we would be talking about formulas differing by a factor that is in the ballpark of a small integer. See ref [52].)

Taking this entropy in a very familiar sense, I shall assume that the macroscopic state of the black hole is thus compatible with $e^{A/4}$ overall internal states. Let us briefly review these matters. We begin with Bekenstein (ref [53], p. 2335):

The connection between entropy and information is well known. The entropy of a system measures one's uncertainty or lack of information about the actual internal configuration of the system. Suppose that all that is known about the internal configuration of a system is that it may be found in any of a number of states with probability $p_n$ for the $n$th state. Then the entropy associated with the system is given by Shannon's formula

$$S = -\sum p_n \ln p_n \qquad \text{[summation over } n\text{; natural logarithm]}$$

This formula is uniquely determined by a few very general requirements which are imposed in order that S have the properties expected of a measure of uncertainty.

Suppose the states are equiprobable; that is, each probability is $1/n$. (I assume that we are in a state of maximal uncertainty about which internal configuration the black hole is in; and by (30) we know that the number of possible configurations is finite.) Then:

(31)     $S = -\sum (1/n) \ln (1/n) = \ln n$

So $n = e^S$. S is how much information we can lack about the internal configuration of the system. A bound on S is thus a bound on how much information there can be about the internal configuration of the system: how many "degrees of freedom"—that is, value-taking variables—the system has. To a factor of $\ln 2$ ($\approx .7$; which we will ignore), this is the same as how many *bits* of information there can be about what is going on in the system. Thus S is a bound on how many bits of information the system can store.

By the Bekenstein-Hawking formula, (30) above, in the case of a black hole we get Susskind's "spherical entropy bound" for free. There can be no more than $A/4$ bits of information about the microphysical states within the black hole. Shockingly, this goes by the *area* of the event horizon, not, as we would expect, by the *volume* it encloses. We would have thought that there could be at least one bit of information for every Planck cube ($\approx 10^{-105}$ m$^3$) of the volume of the black hole. But the real information maximum, the "informational storage capacity" of the black hole, is vastly less than that: no more than on average one bit of information per Planck square ($\approx 10^{-70}$ m$^2$) of the event horizon (less than that, given the "1/4"; but we shall ignore that).

This is assuming that for a black hole the "connection between entropy and information" is the familiar one described above. There are many things one might say about the meaning of the "entropy" of a black hole. For some examples, one might say that (quoted remarks are from ref [54], p. 3):

(i)   "it [is] similar to that of ordinary entropy, i.e. the logarithm of a count of internal black hole states associated with a single black hole exterior"

(ii)  "it [is] the logarithm of the number of ways in which the black hole might be formed"

(iii) "it [is] the logarithm of the number of horizon quantum states"

(iv)  it is the "information lost in the transcendence of the hallowed principle of unitary evolution"

(v)   it is a measure simply of entanglement entropy

(vi)  it is a measure of *renormalized* entropy, subtracting away the entropy of the vacuum fluctuations within the black hole (cf. refs [55]-[58])

But we cannot say (ii)-(vi), or, I think, anything significantly different from (i); for such alternatives ignore the entropic contribution of virtual particle creation well within the black hole, but well away from the black hole's singularity. (It might be suggested that there cannot be vacuum fluctuations within a black hole, because the spacetime there is not well defined. But, when Schwarzschild coordinates are transformed to Kruskal-Szekeres coordinates, the spacetime within the black hole is well defined for regions outside the singularity (ref [12], pp. 555-556; ref [4], pp. 164-167; ref [5], pp. 11-12). There is no reason to hold that the regions where the Kruskal-Szekeres spacelike coordinate $u$ and timelike coordinate $v$ are such that $|u| < v < |(1 + u^2)^{1/2}|$ --that is, within the [future, physical] event horizon but outside the [future, physical] singularity-- are "unphysical." It will become fairly clear below that some of these regions must be physical. Given Heisenberg's uncertainty relations, there must then be vacuum fluctuations.)

Suppose that we *do* ignore the entropic contribution of virtual particle creation within the black hole. Then there *could* be two black holes, otherwise quite comparable (they have the same number of "ways of forming," the same number of horizon quantum states, enshroud the same amount of "lost information," have the same amount of entanglement entropy, and the same renormalized entropy), but differing (at least slightly) in the interior vacuum energy due to virtual particle creation well within the black hole but well away from the singularity. Imagine quiet little regions of "empty space" far within the event horizon and far from the singularity. I am not talking about the conversion near the horizon of virtual particles to real particles through the Hawking process. I am talking about the constant creation and annihilation of virtual particles, well within the black hole, and the resulting vacuum energy. The total negative energy due to virtual fermions will not cancel the total positive energy due to virtual bosons, within the black hole, any more than it does in the universe at large. There will be some net vacuum energy in each black hole, resulting from the vagaries of the constant flux of virtual particles. But then, at least at a given moment, these otherwise comparable black holes could differ in their overall energy, and thus in their overall mass, and thus in their horizon area (which is directly proportional to the square of the mass); and yet still supposedly (by such criteria as (ii)-(vi)) have the same entropy. That is impossible by the Bekenstein-Hawking formula.

In brief, if

(I)  interior vacuum fluctuations do not contribute to the entropy of a black hole

then

(II) it is physically possible that there are two black holes which have the same entropy but differ in (their energy, thus in their mass, thus in the square of their mass, thus in) the area of their event horizons

But (II) is false; hence (I) is false. Hence interior vacuum fluctuations contribute to the entropy of a black hole.

We may then ask: *why* do interior vacuum fluctuations contribute to the entropy of a black hole? It won't do just to answer: "Because they affect the area of the event horizon." That would be like answering the question "What is the meaning of the entropy of a black hole?" with "It is one quarter of the area of the event horizon." The only plausible explanation of why interior vacuum fluctuations contribute to the entropy of a black hole is: "It is because they are part of the internal configuration of the black hole, and *entropy by its very nature* is a measure of the number of distinct internal configurations the black hole can be in." So I shall assume here that the entropy of a black hole is the hidden information about its internal configuration, as in (i). (Note that, if entropy is hidden information, then it is obviously the case that pure states can contribute to this entropy, if they come—as they obviously do, in the case of virtual particle-antiparticle pairs—in different *kinds* which can be *variably arranged* compatibly with the macrostate of the black hole. The additivity of entropy means that the entropy of a set of isolated systems is the sum of the entropies of those systems. It does not obviously follow that the entropy of a system composed of those isolated systems is this sum. Consider a system *X* composed of ten different kinds of isolated zero-entropy systems, which have different arrangements compatible with the macrostate of *X*. If entropy is hidden information, the entropy of *X* is not zero.)

Now, we have seen that in a spatially flat universe (which, again, we are assuming for convenience, since the reality of the things around you cannot turn on the curvature of space) there is no especial difficulty about there being "observable-universe-sized" black holes; indeed, they are statistically inevitable. Suppose, then, that just as you are reading these words your Hubble-luck has run out. (For simplicity, I take the current radius of a Hubble volume to be exactly $10^{26}$ m.) Suppose that a Hubble volume just slightly less than $8 \times 10^{26}$ m away from ours (less by the radius of the Earth) has "gone nonunderdense." It is thus a HBH. We may assume (the reality of the things around you cannot turn on such happenstances) that very little further material ever falls into this black hole in the future history of the universe, so that the radius to the event horizon is not vastly greater than the radius of the HBH: greater by no more than an order of magnitude. So, since the Hubble radius is $10^{26}$ m, let us assume that the radius $r_{\text{EH}}$ to the event horizon, of the black hole the HBH is within, is $10^{27}$ m. Given that assumption (nothing hangs particularly on it; given some other, we need only modify the distance to the HBH), the Earth is then just within the event horizon. You are then living in a black hole.

The Earth, all the furniture of the earth, is within this black hole. In such a large black hole, we being so far away from the singularity toward which our galaxy is now making infall, nothing much is immediately noticeable. You have a maximum of $\pi r_{\text{EH}}/2c$ seconds before you reach the singularity (ref [2], p. 293). But, since $r_{\text{EH}}$ is extremely large ($10^{27}$ m), that is a long time: about $10^{19}$ seconds, or roughly three hundred thousand million years; and it will be a very long time before you have to worry about tidal forces. Everything goes on more-or-less normally for a long time.

But the number of bits of information about what is going on within the black hole is constrained in the way we have described: no more than its horizon area in Planck units, so no more than about $10^{55}\text{m}^2/10^{-70}\text{m}^2 = 10^{125}$ bits of information. That sounds like a lot, but it is not nearly enough. Remember that vacuum fluctuations must count toward the entropy of the black hole, and be included in these $10^{125}$ available bits of information. (Does anyone suppose that a minute after the Earth has passed within the event horizon, of the black hole the HBH is within, while the singularity is still some 138 thousand million light-years away, a physicist would measure some *bare* value of the electron charge, rather than the usual "dressed" value that results from vacuum polarization? Or that, in examining the spectrum of a hydrogen atom, the physicist would find that the Lamb shift had disappeared?) Focus just on virtual photons. The number of joules of vacuum energy from virtual photons (discounting, as is customary, the very highly energetic ones) in one cubic centimeter of space is roughly $10^{116}$ joules (ref [43], p. 75). The average energy of a photon (discounting again the very highly energetic ones) is roughly $10^{-19}$ joules. Thus, in one cubic centimeter of "empty space" there are at any given time roughly $10^{135}$ virtual photons. The available bits of information in our black hole are not enough even to account for the virtual photons in one cubic centimeter of space.

How small are the cubes in our black hole such that there is on average one bit of information about what is going on in each? They are about $10^{81}$ m$^3$/$10^{125}$ = $10^{-44}$ m$^3$. Since there are $10^{38}$ of our little cubes in a cubic centimeter of space, and $10^{135}$ virtual photons, there is thus at any given time about $10^{97}$ virtual photons in one of our little cubes. Thus the informational storage capacity of the black hole, $10^{125}$, is used up simply by ten thousand trillion trillion of our little cubes, occupying a region of only $10^{-16}$ m$^3$; leaving no room for anything else going on.

Thus, once we are within the black hole, the things around you cannot be real. It is physically impossible; they cost too much information, in a perforce informationally impoverished environment. The tables, the walls, the organisms, the planets, the stars, must then be unreal. But if they are unreal *then*, they were *always* unreal. Our Earth's passing under the long shadow of a HBH has not in some magical and instantaneous way suddenly made real things unreal. All goes on normally, for at least a million years. Will you say that the table at one moment was real, and at the next unreal (although absolutely nothing seems to happen to it, and it will remain quite solid and table-like for the next hundred years), simply because it has passed within the boundary of a sphere that is 276 thousand million light-years in diameter? What can an unspeakably distant, and for generations to come harmless, singularity, have to do with the present reality, or unreality, of the table in front of your eyes? It can have nothing to do with it. So the table must never have been real in the first place.

Perhaps you still find HBHs a little strange (though they are statistically inevitable in a spatially flat universe). So, to take a (somewhat) more normal case of a physically possible black hole (but still with flat-space volume calculations), consider again Rindler's hypothetical spherical galaxy, which collapses so that its radius becomes equal to the pseudo-Schwarzschild radius of the mass it contains. It contains $10^{11}$ stars, each idealized to be comparable to the Sun in mass and density, and hence in volume. The radius of the Sun is about 7 X $10^8$ m. Thus its volume is about $10^{27}\,\mathrm{m}^3$. Thus the combined volume of the $10^{11}$ stars is $10^{38}\,\mathrm{m}^3$. The mass of the Sun is about 2 X $10^{30}$ kg. Thus the combined mass of the $10^{11}$ stars is 2 X $10^{41}$ kg. Thus the pseudo-Schwarzschild radius for the mass $M_c$ the galaxy contains (including now ordinary dark matter, e.g. planets, exotic dark matter, and dark energy, as well as the stars) is roughly $10^{15}$ m. Thus the volume of the galaxy, when its radius has shrunk to its pseudo-Schwarzschild radius, and the galaxy is within a black hole, is roughly $10^{45}\,\mathrm{m}^3$. Thus the volume of the galaxy is then $10^7$ times bigger than the combined volume of the stars. "In other words the stars are still far apart when the whole system shrinks within its Schwarzschild radius" (ref [2], p. 290). (By terrestrial standards, "far apart"; the stars would now be separated from each other by on average about 284 solar radii, or about 2 X $10^{11}$ m: 200,000 million meters; less than the distance of Mars from the Sun. This is physically possible for stars, as in binary systems. For example, in the binary system Beta Persei, or Algol, the stars are separated by only about 9,000 million meters.)

Now, if you are worried about the stability of planetary orbits when stars are in such close proximity, or about the likelihood of there being a habitable planet in a spherical galaxy, we could make our point just by talking about stars or rocks. But let us suppose, more colorfully, that an Earthlike planet orbits one of the outermost stars of the spherical galaxy, at a suitable distance for habitability. On that great globe is a city, with palaces, and towers, and temples. In the royal palace, the Princess has just been married to the Philosopher-King, for whom she is now pouring a glass of wine. Just as she is pouring, the galaxy shrinks within its pseudo-Schwarzschild radius, and is within a black hole. We may assume (as before) that very little further material ever falls into this black hole in the future history of the universe, so that the radius to its event horizon is not vastly greater than the pseudo-Schwarzschild radius $r_{PS}$ it has just shrunk within: greater by no more than an order of magnitude. So, since $r_{PS} = 10^{15}$ m, let us assume that the radius $r_{EH}$ to the event horizon is $10^{16}$ m.

As we know, there can be no more bits of information about the microphysical states within the black hole than the area in Planck units of its event horizon. This area is about $10^{33}$ m$^2$, which in Planck units is $10^{33}$m$^2/10^{-70}$m$^2 = 10^{103}$. So there can be no more than $10^{103}$ bits of information about the microphysical states within the black hole. How small are the cubes in our black hole, such that on average each contains one bit of information? They are about $10^{48}$ m$^3/10^{103} = 10^{-55}$ m$^3$.

As we know, given the uncertainty relations there must be virtual particle formation within the galaxy/black hole, at least far from the singularity; for example, in the "empty space" between those stars which themselves are still far from the singularity, say, as the Princess is pouring the wine (or for that matter, in the wine itself, which is mostly empty space). Again, just focus on virtual photons. Since within one cubic centimeter of space there are at any given time something like $10^{135}$ virtual photons, and there are $10^{49}$ of our little cubes in a cubic centimeter of space, there is thus at any given time about $10^{86}$ virtual photons in one of our cubes, and so (at least) $10^{86}$ bits of information used up. Remember that the informational storage capacity of our black hole is only $10^{103}$. So this is used up by simply a hundred thousand trillion of our little cubes (just *counting* virtual *photons*), occupying a region of only $10^{-38}$ m³ of the empty space between the stars; leaving no room for anything else going on within the black hole.

Therefore, when her galaxy is become a black hole, the Princess's body, and the wine, and the glass and the table it is set upon, cannot be real. It is physically impossible. (They cost too much information, in an informationally impoverished region of space; in the information-desert that is a black hole.) The cloud-capp'd towers, the gorgeous palaces, the solemn temples, the great globe itself, must all be insubstantial. But if they are unreal *then*, they were always unreal. The galaxy's shrinking within its pseudo-Schwarzschild radius cannot in some instantaneous way make real things unreal. All goes on normally, at least for a little while. The Wedding guests would not immediately notice that their galaxy is become a black hole. They have a maximum of about a year and a half, before they reach the singularity; and tidal forces will very soon be felt. (Notice that their galaxy is now only about one-fifth of a light-year in diameter.) But the table the glass sits upon would scarcely immediately cease to exist; if it had ever existed at all.

Now, if those tables and walls, etc., were always unreal, then so are the ones around you. There can be no physical distinction about reality between them, based on whether they reside in an unlucky galaxy. The Princess's body is like your body, her table like your table. It is all earthy stuff, like in kind, and "one touch of Nature makes the whole world kin." These things we see or touch cannot be real. This world is an illusion; a vast, long-lasting, law-governed illusion.

So now let us give our argument, not by appeal to HBHs or Rindler's galaxy, but with "ordinary" supermassive black holes, of the sort at the heart of galaxies. Physics tells us this:

(i)   If this table exists then it (in its local reference frame) could enter a supermassive black hole

Put it in a spaceship and send it into the black hole! (A subtlety: in our reference frame, it would never reach the horizon of the black hole; we would see it infinitely slow down. But in the local reference frame of the spaceship, and of the table, it would pass easily into the black hole.) Physics also tells us this:

(ii)  If this table exists then if it (in its local reference frame) were to enter a supermassive black hole then it would still exist and begin infall to the singularity

Upon crossing the event horizon the table cannot simply, literally, *cease to exist*. If it does, its mass/energy is lost, being available neither to the black hole nor to the universe outside the black hole, which violates conservation of mass/energy. (In fact, for a sufficiently large black hole, the table will not even immediately be "spaghettified" by tidal forces; nothing special happens to it immediately. In any event, it cannot simply, literally, cease to exist.) Physics also tells us this:

(iii) If this table exists then if it (in its local reference frame) were to enter a supermassive black hole then it would not exist

For there is not enough informational storage capacity, even within a supermassive black hole, for such things as the mass/energy of a table. The spacetime within the black hole (but outside the singularity) is well defined in Kruskal-Szekeres coordinates. There is "empty space," as between the stars in Rindler's hypothetical galaxy. The uncertainty relations are inviolable: there must be vacuum fluctuations. These must be counted among the available bits of information there can be, about the microphysical states of the black hole. Their number is enormous.

So suppose, for a conditional argument:

(iv) This table exists

By that and (i) we have it that it is physically possible for it (in its local reference frame) to enter a supermassive black hole, and since whatever is physically possible is possible:

(v)  Possibly, it (in its local reference frame) enters a supermassive black hole

Given (v), the counterfactuals which are the consequents of (ii) and (iii) are not both true. (Two counterfactuals, themselves having contradictory consequents, are both true only if they have an impossible antecedent, which, by (v), they don't.) Thus, given (v) we have:

(vi)  Either it is not the case that [if it (in its local reference frame) were to enter a supermassive black hole then it would still exist and begin infall to the singularity] or it is not the case that [if it (in its local reference frame) were to enter a supermassive black hole then it would not exist]

From (ii), (iii), (vi) it follows that:

(vii) It is not the case that this table exists

So by our conditional argument we conclude:

(viii) If this table exists then it is not the case that this table exists

Therefore this table does not exist.

        And, of course, this is not just a peculiarity of the table. The point holds for all the things around you. The lovely lakes and meadows of Mount Assiniboine in British Columbia; the cold, dry deserts of Mars; the selenium snow that falls like golden glitter in the clouds of Venus; the warm ocean within Europa; the methane seas of orange-misted

Titan; the frozen "Earth" that orbits a red dwarf star twenty thousand light-years inward in the Milky Way; all these things are without substance.

But, in a sense, they are there. You could go and look at them. Our non-existent Sun (Helios in his golden chariot) rests between two arms of the Milky Way; and the Milky Way, in a local group of galaxies, resides at the utmost fringe of the Virgo supercluster. Go on much farther in that direction, passed the Virgin. To your right is the Coma supercluster of galaxies, to your left the Centaurus. Ahead is the Shapley supercluster, and then the great Bootes void. Then the Bootes supercluster, with its many worlds. At that point you're about a thousand million light-years from the Earth. All these things are insubstantial; mere semblances of being. Physics teaches us that they cannot be real.

One avenue of escape would be to rewrite physics somehow. But I prefer to keep the well-established laws of physics, and simply believe what physics is telling us: that the universe around us is not real. This universe is a vast, long-lasting illusion; but an illusion such that, if we pretend that its objects are real, and calculate their behavior in the manner in which physics instructs us to do, we get extremely accurate predictions about how the illusion will continue: that is, what our meter readings and other observations will be. (Quantum electrodynamics, for example, is able to make predictions accurate to 29 places of the decimal.) Consider a book on stellar evolution. It should not be thought of as being committed to the metaphysical claim that there are stars. Rather, it says: if there are stars, then this is how they behave; calculate accordingly, pretending there are stars, and you will make accurate predictions. Why should a physicist venture into metaphysics? Don't ask the "Thales question": What is there? (Answer: very little, at least around us.) Ask the "Heisenberg question": What formalism should we adopt in order to make accurate predictions?

Though the stars, and the Earth, and the supposed objects therein are not real (they can't be; it is physically impossible; such things cost too much information), each of us should be able to convince ourselves, in a Cartesian way, that we are real. (*Cogito ergo sum. Je pense, donc je suis.* I think, therefore I am.) We ourselves may indeed be physically real objects (but not our supposed "bodies"; they are unreal; they cost too much information, and in any event you can't make them without stars). And there may indeed be (I should say that there surely is) a physically real world; it just isn't *this*. Anything you see or touch, in the ordinary way of things, is unreal. It is only the things which are normally unseen that are real: you, and I, and numbers, and perhaps gods and physically real heavenly realms.

Oh, but surely there are gods and physically real heavenly realms! *We* are real, the world around us is not; physics teaches us that it is made of ghosts. Surely there must be a world that is physically real, and our natural home. Surely that is what men call heaven; a paradise some claim to have visited. The generic skeptical response to such reports (e.g., to near-death experiences) is: *something weird is happening in the brain*. But there are no brains; or organisms, or planets, or stars. All that is illusion; a metaphysical myth passed down to us from early Victorian times, of a billiard-ball universe spotted with protoplasm. How heavy and immovable the universe was for them. Does the sheer weight of the material world impress you? It can be no heavier than the corresponding black hole—in which there is a great emptiness, and lightness of being.

It seems to me that we should then take reports of heavenly encounters, many of which are wholly credible, with great seriousness, focusing, like good empiricists, on what is clear and common. They speak of a world vaster and more colorful than this, solid and real, and bright like a diamond struck by sunlight. Those who come back can scarcely stop telling us how very real that other place is; a many-colored land, where your mind and senses are clear and keen, a land of summer breezes and the greenness of fields. Under a sky that is unimaginably high, there is a vast Earthlike landscape there, of which everything you ever saw in this world that seemed to you beautiful was but the palest copy. There are cities there. Our loved ones are there. God is there.

Don't you remember, when you were very young, the sense of wonder that came over you when you first read the Greek myths? The sunlit Greeks, the temples and the sea, the summer storms the cloud-gatherer brings, the sense of color and wonder? It was the heroism and adventure (as much intellectual as physical) that attracted you to the stories; not the dreariness of vengeance and battles and smoking cities. It was the sense of an infinite unknown world that contains both men and gods; Apollo and Euclid, geometry and theology carved out of sunlight. It is all true, in the real world, just on the other side, when you step out of the illusion. There are *real* hands, and faces, and woods and picnics, and voyages and theorems and creatures and worlds beyond reckoning. And there is heroism and adventure, as He who sits upon the throne knows right well; who gives to real and right and solid beings an eternal weight of glory.